\pdfoutput=1
\documentclass[preprints,article,accept,pdftex,moreauthors]{Definitions/mdpi} 
\usepackage{subfig}
\usepackage{underscore}
\firstpage{1} 
\pubvolume{1}
\issuenum{1}
\articlenumber{0}
\pubyear{2025}
\copyrightyear{2025}
\datereceived{ } 
\daterevised{ } % Comment out if no revised date
\dateaccepted{ } 
\datepublished{ } 
\hreflink{https://doi.org/} % If needed use \linebreak
\Title{Superfluid He-4 as Dark Matter Detector}

\TitleCitation{Superfluid He-4 as Dark Matter Detectors}

\Author{ Zizhou Huang $^{1}$\orcidD{}, Mihaela Parvu $^{2,3,*}$\orcidB{}, Dmitry Budker $^{4,5,6}$\orcidA{}, Ariel Zhitnitsky$^{7}$\orcidE{} and Konstantin Zioutas $^{8}$\orcidC{}}

\AuthorNames{Dmitry Budker, Zizhou Huang, Mihaela Parvu and Konstantin Zioutas}

\isAPAStyle{%
       \AuthorCitation{Budker, D., Parvu, M., Zioutas, K. \& Huang, Z.}
         }{%
        \isChicagoStyle{%
        \AuthorCitation{Budker, D., Parvu, M., Zioutas, K. and Huang, Z..}
        }{
        \AuthorCitation{Budker, D.; Parvu, M.; Zioutas, K.; Huang, Z.}
        }
}

\address{%
$^{1}$ \quad Department of Physics, University of Science and Technology of China, 96 Jinzhai Road, Hefei, Anhui 230026, China; 
hzz030830@mail.ustc.edu.cn\\
$^{2}$ \quad University of Bucharest, Faculty of Physics, POBox MG-11, Bucharest-Magurele, Romania; mihaela.parvu@unibuc.ro\\
$^{3}$ \quad Institute of Space Science — INFLPR Subsidiary, 077125, Magurele, Romania\\
$^{4}$ \quad Helmholtz-Institut Mainz, 55128 Mainz, Germany; budker@uni-mainz.de\\
$^{5}$ \quad Johannes Gutenberg-Universit{\"a}t Mainz, 55128 Mainz, Germany\\
$^{6}$ \quad Department of Physics, University of California, 94720-7300, Berkeley, USA\\
$^{7}$ \quad Department of Physics and Astronomy, University of British Columbia, Vancouver, Canada; arz@phas.ubc.ca\\
$^{8}$ \quad 
Physics Department, University of Patras, GR 26504 Patras-Rio, Greece; Konstantin.Zioutas@cern.ch}

\corres{Correspondence: mihaela.parvu@unibuc.ro;} 
\abstract{Superfluid helium-4 is an attractive medium for high-sensitivity detection of dark matter (DM) candidates (as well as other particles). There are several collaborations that have built and tested superfluid helium detectors, particularly focusing on the sub-GeV DM mass range. In this paper we discuss a novel idea for detecting Axion (Anti)Quark Nuggets (AQN) by using parasitically the superfluid helium-4 LHC cooling system as a unique large-scale earth bound DM detector. In the case of the AQN, we address the question of propagation in the metal parts surrounding the helium. An AQN induces significant heating around the propagation track that damages the enclosure.}


\begin{document}

\section{Introduction}

It has been proposed previously that anti-quark nuggets (AQNs) could be detected via their interactions with the LHC beam \cite{zioutas2025aqn}. The challenge lies in the limited active surface area available for such reactions. With an expected AQN flux of approximately 0.04\,$(10^{25}/B)$\,/\,km$^2$ / year, where $B$ is the typical baryon charge of the AQN \cite{Zhitnitsky:2002qa}, the increase of the active surface area is the solution.

In this proposal, we present a novel approach that significantly increases the sensitive area, thereby enhancing the probability of detecting AQNs. We consider the interactions between AQNs and the 27 km long LHC cryogenic system, the temperature of which is carefully monitored as required to prevent superconducting magnet quenches. This system provides a unique opportunity to explore AQN interactions due to its extensive coverage and built-in advanced monitoring capabilities.

Superfluid helium-4 (SF He-4) is an excellent detector material in the search for dark matter (DM). When a particle interacts with SF He-4, the energy release triggers a series of processes, distributing the deposited energy among quasiparticles, light (mainly ultraviolet with some infrared), stable triplet excimers, etc.

\begin{figure}[ht]
    \centering
    \includegraphics[width=.9\linewidth]{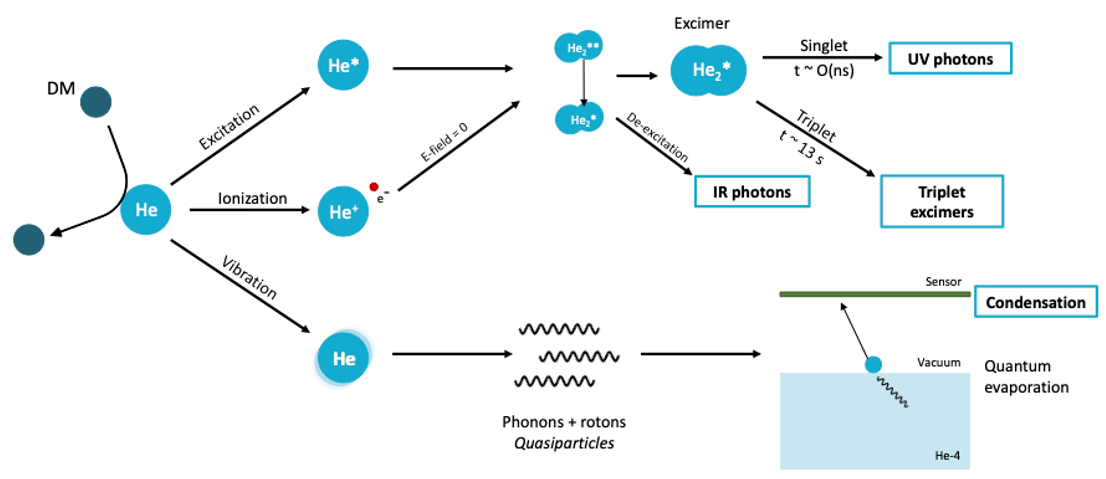}
    \caption{Schematic illustration of the microscopic signal-generation processes in superfluid $^4$He following an interaction of an incident dark-matter particle. Atomic excitation produces excited helium atoms that form excimers, whose singlet and triplet states give rise to ultraviolet photons and long-lived triplet excimers, respectively. Ionization produces electron--ion pairs that may recombine in the absence of an electric field to form excimers or undergo de-excitation accompanied by infrared photon emission. Energy transferred to atomic motion generates phonons and rotons. Adapted from Ref.~\cite{DELightExperiment}.}
    \label{SFHe-4_detection-principle}
\end{figure}

The distribution of the deposited energy among different channels depends on the interaction mode, which signifies, for example, that the incident particle interacted with nuclei and/or the electrons \cite{DELightExperiment}. A schematic view of the interaction mechanism and detection principle of SF He-4 can be seen in Figure \ref{SFHe-4_detection-principle}.

There are various experimental ideas for utilizing SF He-4 and SF He-3 for detecting neutrinos, dark matter, gravitational waves etc. Some authors proposed using SF helium to detect DM axions via magnetically induced couplings. They explore the use of the homogeneous precession domain (HPD) of SF He-3 to detect axion dark matter. The axion wind could cause a small shift in the precession frequency of a large-amplitude NMR signal in SF He-3 \cite{Foster_2024}.

SF He-4 is being actively investigated as a medium for detecting DM particles in the sub-GeV mass range. Several experimental concepts and prototypes have been developed to leverage the unique properties of SF He-4 for particle detection. For example, the HeRALD collaboration \cite{Anthony_Petersen_2024} is aiming to detect light DM particles; HeLIOS \cite{Hirschel_2024} plans to observe ``new'' forces mediated by ultralight DM particles; DELight \cite{10.21468/SciPostPhysProc.12.016} investigates sub-GeV dark matter interactions; HERON \cite{doi:10.1142/9789812778000_0008} focuses on low-energy neutrino detection. The detector volumes in these experiments range from $\sim10^2$ to $\sim10^5$\,cm$^3$. By comparison, the volume of SF He-4 of the cryogenic ring of the LHC is $400$\,m$^3$. It is this comparison that makes the LHC system so attractive for rare DM events. In the following we present a possible detection principle using the LHC SF He-4 cooling system as a detector for rare DM events like the axion antiquark nuggets (AQNs) proposed in 2003 by A.\,Zhitnitsky \cite{Zhitnitsky:2002qa}, see a brief review \cite{Zhitnitsky:2021iwg} for a general introduction and motivation of the AQN model. In particular, it explains why antimatter nuggets (which play a key role in this work) appear  naturally in the AQN framework as a result of resolution of two  fundamental problems in cosmology: a) why the baryonic and the DM components are  similar in values; b) why there ere nearly no visible antibaryons in the Universe. 

\begin{table}[h]
    \centering
    \begin{tabular}{|l|c|c|}
        \hline
        \textbf{Experiment} & \textbf{Current SF He Volume} & \textbf{Future Development Volume} \\
        \hline
        \textbf{HeRALD}  & $\sim$10 g ($\sim$0.01 L) & 10–1000 g ($\sim$0.01–1 L) \\
        \hline
        \textbf{HeLIOS}  & 145 ml & Scalable for improved sensitivity \\
        \hline
        \textbf{DELight} & 10 L & $\sim$100 L \\
        \hline
        \textbf{HERON}  & Small-scale prototype ($\sim$liters) & Multi-ton scale planned \\
        \hline
        \textbf{LHC cryosystem} & $4 \times 10^5$ L & FCC ($\sim 10^6$L) \\
        \hline
    \end{tabular}
   \caption{Current and future superfluid helium volumes in various experiments.}
    \label{SFHe_experiments}
\end{table}

\section{The large scale LHC SF He-4 system}

\begin{figure}[ht]
    \centering
    \includegraphics[width=0.75\linewidth]{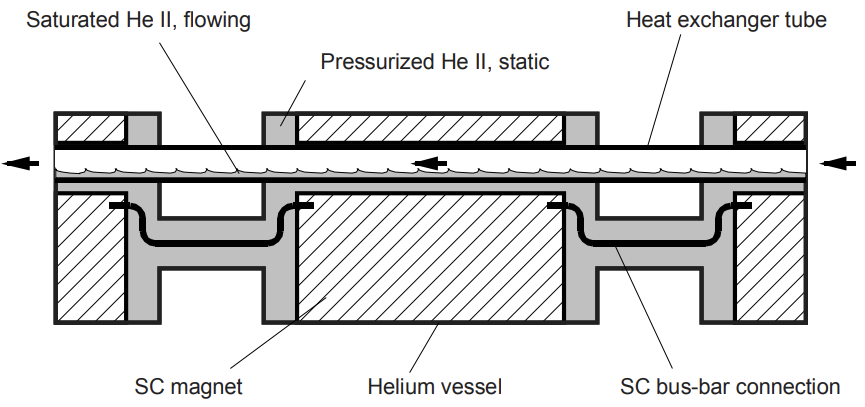}
    \caption{Principle of the LHC superfluid helium cooling scheme. The superconducting magnets are immersed in static baths of pressurized SF He-4, which transport the deposited heat by thermal conduction to a copper heat-exchanger tube extending along the magnet string. Inside this tube, flowing saturated two-phase SF He-4 absorbs the heat continuously through evaporation, providing a distributed quasi-isothermal heat sink while maintaining hydraulic separation between the pressurized magnet baths and the cooling flow. Adapted from Ref.\,\cite{Lebrun_SFHE-4}.}
    \label{SFHe-4_cooling-scheme}
\end{figure}

The superconducting magnetic system of the LHC is cooled by pressurized helium-4 at 1.9\,K and 0.1\,MPa, with heat transported via conduction to a heat-exchanger tube along the magnet string. Inside this tube, saturated helium II (the superfluid phase) absorbs the heat load through gradual evaporation, providing an isothermal cold source with a large exchange area, allowing heat extraction with minimal temperature difference. This design ensures hydraulic separation between the magnet baths and the cooling loop, preventing contamination and inhibiting the propagation of quenching \cite{LEBRUN19941}. A schematic view of this system is presented in Fig.~\ref{SFHe-4_cooling-scheme}. The SF He-4 layer has a minimum thickness of $\sim1.6$\,cm in most parts of the LHC ring. The cross-section area of the LHC SF He-4 vessel is diameter $\times$ circumference = 27.6 $\times 10^{-5}$\,km$\times$ 27\,km\,$\approx 7.45 \times 10^{-3}$\,km$^2$. The LHC cooling system of about 400 m$^3$ SF He-4 has a mass of 58\,t \cite{Brüning:782076}. For comparison, the He-4 mass required for the cooling system of the future circular collider (FCC) is estimated at 230\,tonnes.

\section{AQN detection principle and numerical results}
\label{sec:AQN_det+principle}

% Given the interatomic distance between He-4 atoms of approximately 2.2 Å, an AQN with a radius of approximately $10^{-5}$ cm would interact simultaneously with $\sim 10^8$ helium atoms.
When AQNs interact with matter, protons can penetrate the AQN core and undergo annihilation near the surface. This process releases an energy of approximately 2\,GeV per annihilation. This energy is then transferred to positrons from the AQN electrosphere, which acquire 2 to 5\,MeV and propagate out of the surface of the nuggets \cite{Forbes_2008}. These positrons are accelerated by the large electric fields of the AQN electrosphere, releasing their energy as Bremsstrahlung radiation. The properties of the electrosphere determine the scale and spectral characteristics of this emission, where the keV scale and the approximately ``thermal'' nature of the spectrum arise naturally \cite{Forbes_2008}. The spectral surface emissivity of a nugget is shown in Fig.\,2 of Ref.\,\cite{Ge_2022}.  

The energy radiated per unit AQN path length by the heated up AQN is given by  \cite{Budker_2022}: 
\begin{equation}
    -\frac{dE}{dx} 
 \approx \kappa \cdot \pi ~R^2 \cdot 2 ~\mathrm{GeV} 
 ~\cdot ~n\,,
\end{equation}
where $n$  is the number density of nucleons such that the density is $\rho\approx n~m_p$ and $\kappa$ represents a dimensionless parameter that accounts for the fact that not all matter impacting the nugget undergoes annihilation and not all the energy released by an annihilation event is fully thermalized within the nuggets. For instance, a portion of the energy can be emitted in the form of axions and neutrinos.

Considering $\kappa \approx 1$ expected for the atmosphere, the energy released is $\sim 8$\,kJ/m. For the case of higher-density  environment such as solid or liquid, low-energy positrons can be stripped from the electrosphere and the suppression factor is $\kappa \sim 10^{-2}$ or smaller \cite{Budker_2022}. 

When the AQN propagates in a dense medium, the heat from the electrosphere surrounding its core is transferred to the core of the AQN, accompanied by the emission of photons with energies up to some 100\,keV \cite{Ge_2022}. For an AQN traversing the LHC cryo-ring (0.8\,m of iron), the total energy release is approximately $3 \times 10^5$\,J, 
%\textbf{How is this obtained? What length and in which material?} 
corresponding to $10^{19}$ of $\sim$ 100\,keV photons, which is to be compared to $\sim 10\,$MeV/cm energy deposition by a relativistic charged particle. Even though only a fraction of the emitted photons deposit their energy within the SF He-4, this is still sufficient for detection. 

A question is frequently asked: if the energy of $3 \times 10^5$\,J is released over less than a meter of the AQN passage through the iron, would this not produce a catastrophic effect? 
% The answer to this is that the energy is mostly released in the form of x-rays with a characteristic energy between 100 and 200\,keV, which spreads the heat over a ``tube'' of a characteristic diameter on the order of attenuation length of the x-rays, which is on the order of 1\,cm. \textcolor{green}{We estimate that the iron ``tube'' surrounding the AQN track heats up by about 400\,K and cools off due to heat exchange on a time scale of 1\,s. While such an event causes significant transient strain in the material, it is not clear if this will cause any detectable damage. While this requires further detailed materials-science analysis, we assume that nothing dramatic happens to the material.} A
Another concern may be the actual ``track'' of the AQN on the order of 10$^{-5}$\,cm in diameter, where a significant fraction of the nucleons are annihilated upon the AQN passage. One may wonder if this would, for example, break the vacuum tightness when an AQN passes through a wall of a vacuum vessel or whether a passage of an AQN through the walls of a superfluid vessel would cause a superfluid leak.

These concerns are discussed in Sec.\,\ref{Sec:AQN+prop_metal}.
% Again, this is an issue that needs to be carefully studied; \textcolor{green}{for now, we assume that the material melts locally (on the scale of 10$^{-5}$\,cm) so that it ``self-repairs'' and these hypothetical leaks do not occur.}

\section{Monte Carlo simulation}

\begin{figure}[ht]
\centering
\subfloat[2D view with component names. \label{fig:sub1}]{%
   \includegraphics[width=0.48\linewidth]{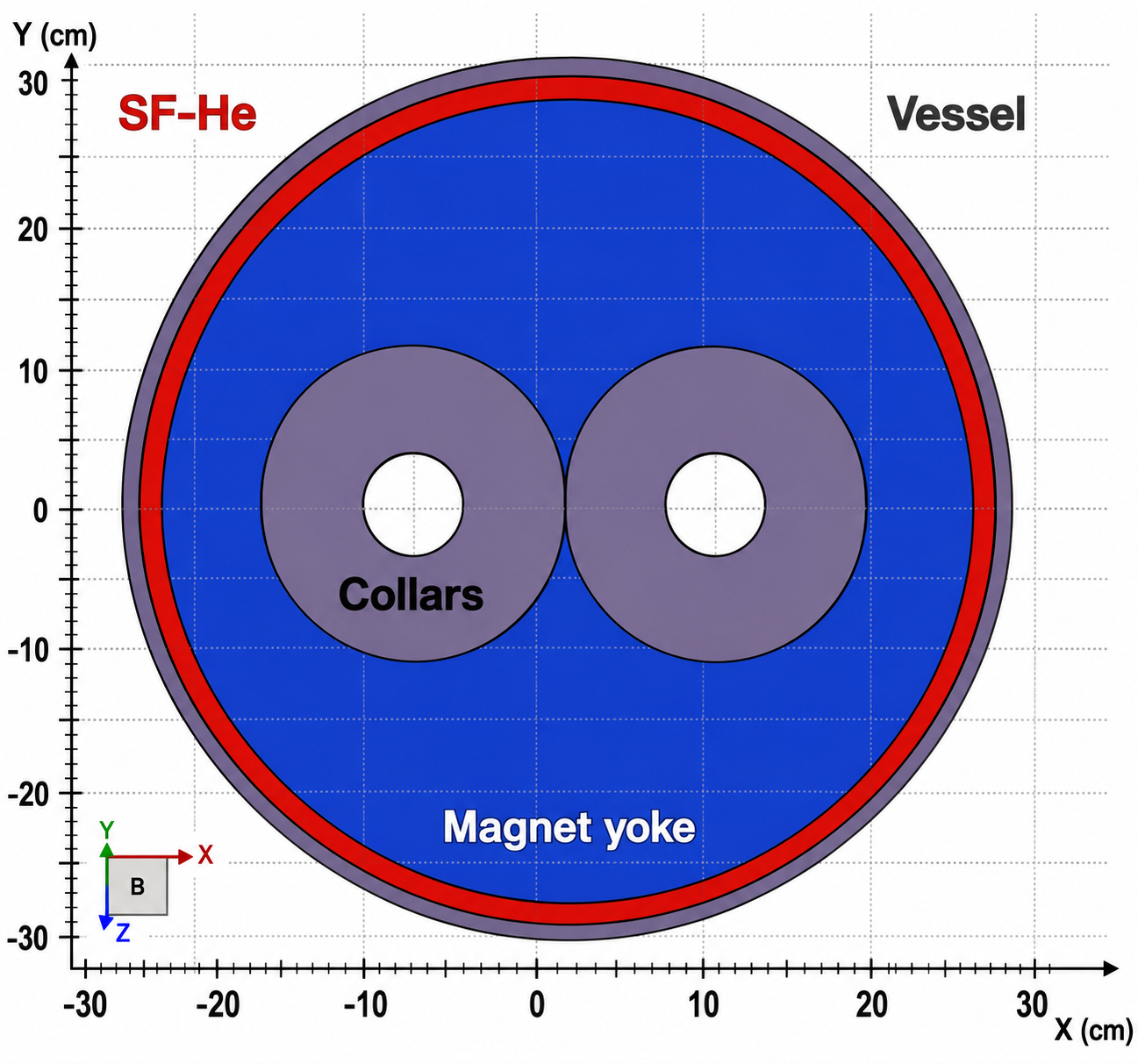}}
\hfill
\subfloat[3D view of the setup. \label{fig:sub2}]{%
   \includegraphics[width=0.48\linewidth]{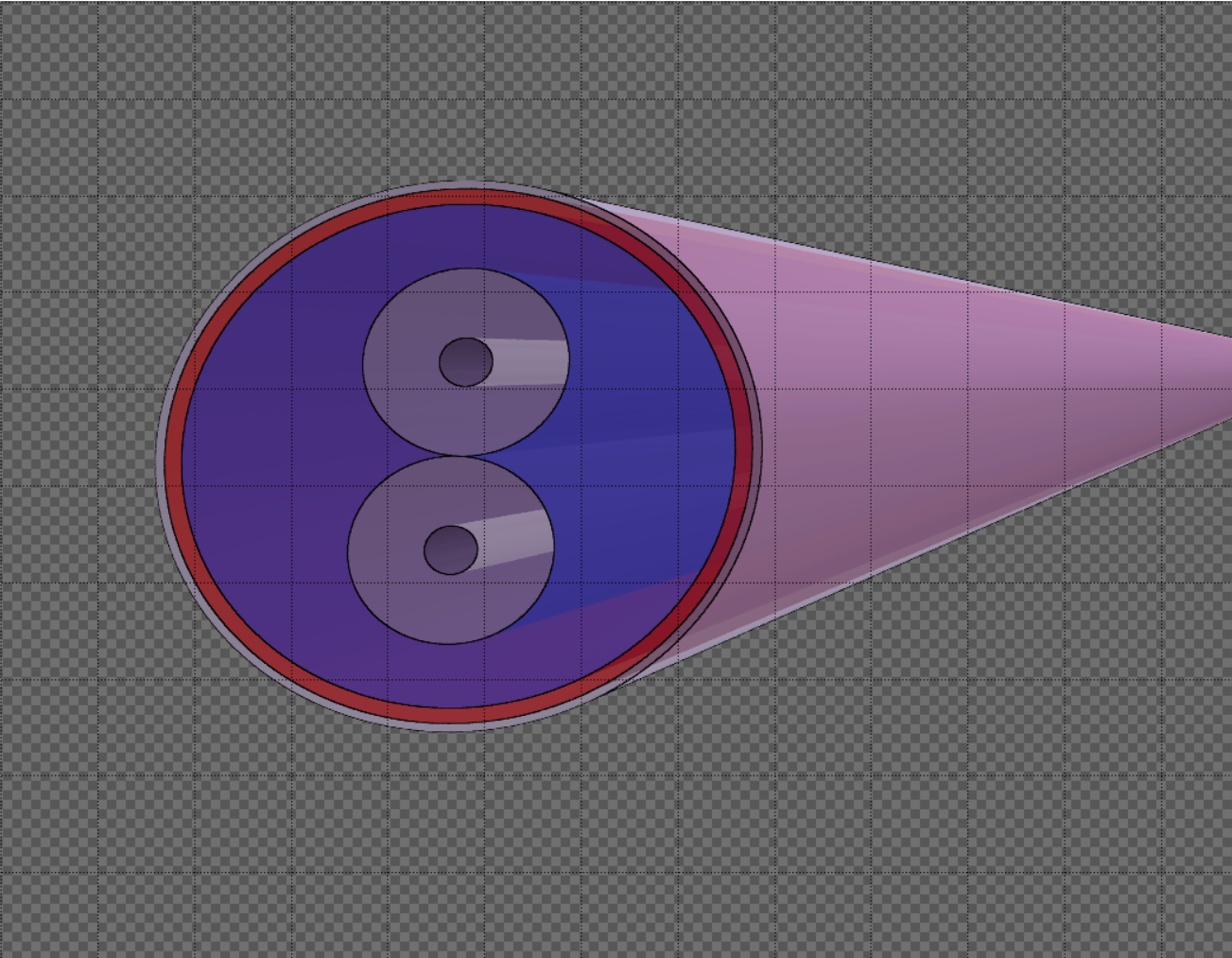}}
\caption{Geometry of the simplified LHC cryogenic system implemented in the FLUKA simulations. (a) Cross-sectional view identifying the principal structural components... (b) Three-dimensional rendering...}
\label{fig:FLUKA-geometry}
\end{figure}

To estimate the energy deposition and the temperature increase due to the interaction of AQNs with the SF He-4 we performed Monte Carlo simulations using FLUKA \cite{10.3389/fphy.2021.788253} and GEANT4 \cite{ALLISON2016186}. We implemented a simplified but realistic geometry that reflects the essential structure of the LHC cryogenic system (see Fig.\,\ref{fig:FLUKA-geometry}). The outer boundary is defined by a stainless-steel (SS316LN) cylindrical shell with a radius of 28.5\,cm, representing the shrinking cylinder that surrounds the SF He-4 layer. Within this shell, there is a cylindrical volume of superfluid helium (SF He-4) with a radius of 27.6\,cm, which acts as the cooling medium. Further inside, the magnet yoke is modeled as a solid cylinder with a radius of 26.0 cm, composed of low-carbon silicon steel. Additionally, two stainless-steel cylindrical collars are positioned symmetrically along the x-axis at $\pm 9.7$ cm, each having a radius of 9.7\,cm.

% \textcolor{red}{names of the regions bigger and with other colors.}

For 100\,keV photons, the attenuation length in SF He-4 is $\sim$ 0.46\,m, significantly greater than the $3.4 \times 10^{-3}$\,m attenuation length in iron. With the density of SF He-4 of 0.145\,g/cm$^3$ and accounting for the metallic components (such as 316LN, low-carbon steel, and YUS 130S \cite{Brüning:782076}), Monte Carlo simulations indicate that the energy absorbed by SF He-4 due to an incident AQN is significant: $\sim$60\,J/cm.

Considering a cylindrical geometry with a radius of 47\,cm for SF He-4 for the metallic components, an AQN traversing the tube of the LHC cryogenic system along its diameter could increase the SF He-4 temperature from its operating level of 1.9\,K by approximately 4.4\,mK, if the energy is absorbed uniformly by the most conservative estimate, if energy is immediately absorbed uniformly in SF He-4. Given the thermometer resolution of 1\,mK, such a temperature increase can be detected. The real temperature rise will be more significant due to concentration of energy near the AQN trace, as is stated afterwards. Concurrently, the temperature of the metallic components could rise to several hundred kelvins. Since the superfluid transition temperature of SF He-4 is 2.17\,K, exceeding this threshold would induce a local phase change, reverting superfluid helium to its normal liquid state. Such phase transitions might serve as an initial indication of an AQN event and would be subjected to rigorous verification to confirm their origin.

The heat absorbed by the metallic components of the LHC ring would predominantly be transferred to the SF He-4, resulting in a quasi-prompt temperature change in the helium layer. With a heat conduction wave velocity of approximately 20 m/s \cite{Vinen:808382}, the temperature of SF He-4 dissipates rapidly, initiating a subsequent decrease. 

A COMSOL \cite{COMSOL} finite element analysis was conducted to simulate the thermal conduction process subsequent to an AQN interaction within a section of the LHC cryogenic system. The length of this section is maximally 0.8\,m when an AQN hits at the center of the cylinder and goes along its diameter. The model included an outer vessel made of 316LN stainless steel, an intermediate layer of liquid helium-4, and an inner core of pure iron, closely replicating the actual geometry of the LHC pipeline. It was initially assumed that an AQN would impact perpendicularly at the midpoint of the segment, creating localized heating zones around the AQN track with radii on the order of 1\,cm in the metallic components and 10\,cm in the helium layer, as estimated based on the X-ray absorption lengths. The initial temperature profile exhibited a sharp gradient, increasing from the nominal operating temperature of 1.9\,K to values surpassing 2.36\,K in the helium layer, while the metallic regions experienced brief heating up to a thousand kelvin. The thermal conductivity of superfluid helium-4 was estimated based on simulations aligning with the observed heat-wave propagation speed of 20\,m/s, whereas material properties for stainless steel and iron were sourced from COMSOL’s built-in databases and references \cite{Vinen:808382,10.1063/1.556028,BUCKINGHAM196180}. Over a simulated duration ranging from 0 to 0.02\,s, the results showed that the helium layer sustained temperatures above 2.0\,K in localized regions.

\section{AQN propagation in metal}
\label{Sec:AQN+prop_metal}

In this section we analyze in more detail the thermal response of the metallic components of the LHC cryogenic system to the passage of an AQN. In a commonly used simplified picture, the released energy is assumed to be distributed uniformly over a cylinder with a diameter comparable to the photon attenuation length; however, this is an approximation. In fact, the energy deposition is strongly peaked near the AQN path, leading to the formation of an extremely hot central core.

We consider an AQN incident perpendicularly on the cylindrical metallic structure of the LHC. The AQN path is approximated as a thin straight line. Antiproton annihilations occur along this line, and the resulting secondary X-rays (see Sec.\,\ref{sec:AQN_det+principle}) are emitted approximately uniformly from the track.

After averaging over the longitudinal emission angle, the problem can be reduced to an effective two-dimensional model in the transverse plane. The AQN track becomes a point source at the origin, and energy deposition is described in terms of the radial coordinate $r$.

Two assumptions are introduced for the energy absorption:

\begin{enumerate}
    \item The absorbed energy does not depend on the angle variable but does depend on the radial distance $r$ from the AQN track. The energy absorbed between $r$ and $r+\mathrm{d}r$ is uniformly distributed over the circular ring of area $2\pi r\,\mathrm{d}r$.
    \item The cumulative absorbed energy inside radius $r$ follows an exponential attenuation law,
    \begin{equation}
        E(r) = E_0\left(1 - e^{-r/R}\right)\,,
    \end{equation}
    where $R$ is the characteristic photon absorption length in the material.
\end{enumerate}

Differentiating, we get

\begin{equation}
    \mathrm{d}E(r) = \frac{E_0}{R} e^{-r/R} \, \mathrm{d}r\,.
\end{equation}

Dividing by the ring area $2\pi r\,\mathrm{d}r$ gives the local absorbed energy density

\begin{equation}
   \varepsilon (r) = \frac{E_0}{2\pi R r} e^{-r/R}\,. \label{Eq:Eps(r)}
\end{equation}

\begin{figure}
    \centering
    \includegraphics[width=1.0\linewidth]{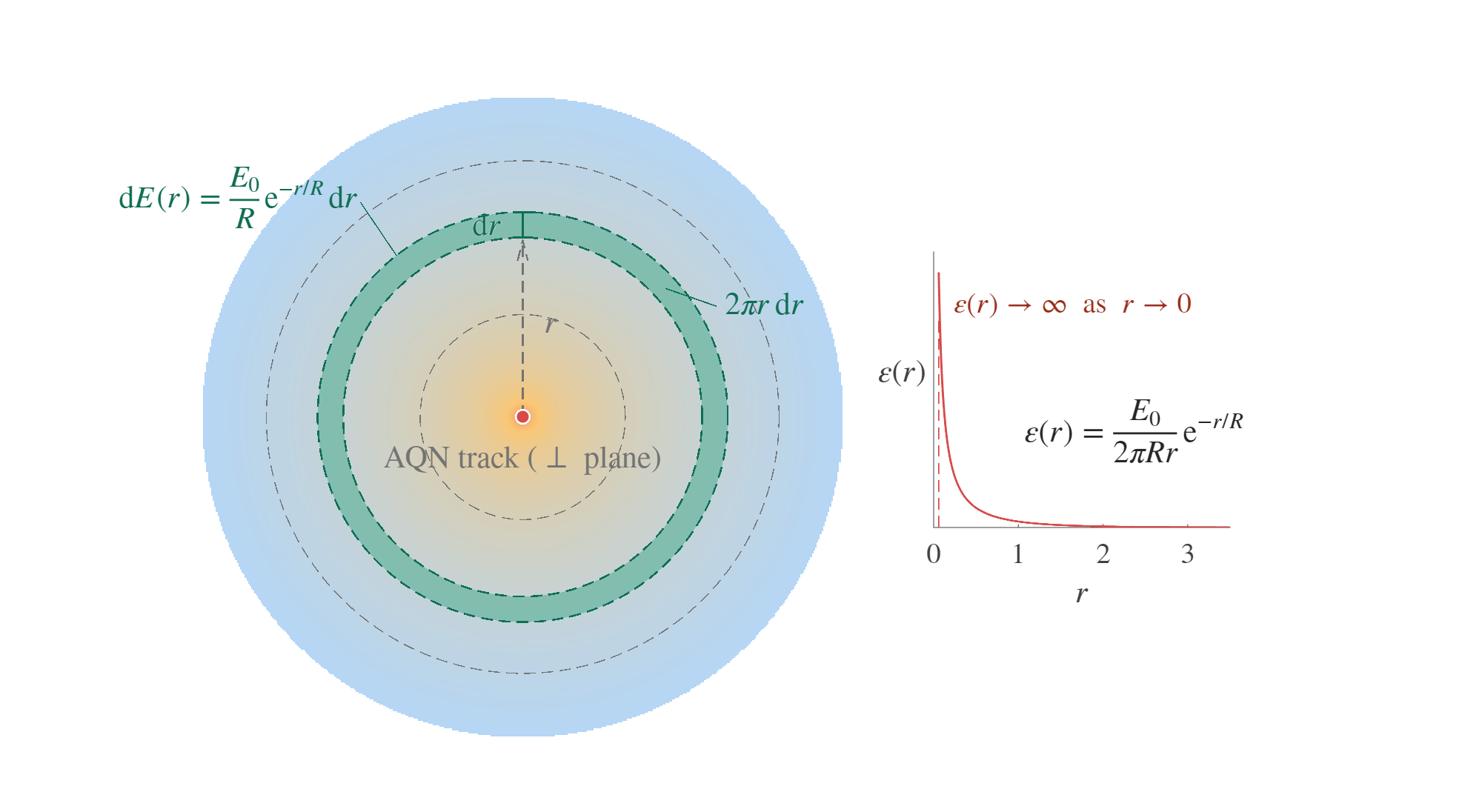}
    \caption{Schematic illustration of the effective two-dimensional model used to describe energy deposition by an axion quark nugget (AQN) traversing the metallic components of the LHC cryogenic system. After averaging over the longitudinal direction of the AQN trajectory, the track is represented as a point source at the origin. The deposited energy between radii $r$ and $r+\mathrm{d}r$ is assumed to be uniformly distributed over the annular area $2\pi r\,\mathrm{d}r$, resulting in the radial energy-density distribution $\varepsilon(r)$ shown on the right. This analytical model provides the initial temperature profile used in the subsequent thermal simulations.}
    \label{fig:2Ddiagram}
\end{figure}

A schematic illustration of this model and the resulting energy density profile is given in Fig.~\ref{fig:2Ddiagram}.

Let $C(T)$ denote the specific heat capacity and $\rho(T)$ the mass density of the material. The absorbed energy at radius $r$ heats the material from the initial temperature $T_0$ to a final temperature $T_{\mathrm{end}}(r)$:

\begin{equation}
       \varepsilon (r)
    =
    \int_{T_0}^{T_{\mathrm{end}}(r)} C(T)\,\rho(T)\,\mathrm{d}T\,. \label{Eq:epsilon(r)}
\end{equation}

A crucial feature of Eq.\,\eqref{Eq:Eps(r)} is that the left-hand side, i.e., $\varepsilon(r)$, diverges as $r \to 0$ because of the $1/r$ factor. Consequently, the upper limit of the temperature integral must formally diverge. In other words, close to the AQN path the predicted temperature rise becomes unbounded. In fact, we restrict the consideration to radii larger than the size of the AQN itself taken here to be $\sim 0.1\,\mu\mathrm{m}$. 

% must cut the expression off at the size of an AQN, taken here to be $\sim 0.1\,\mu\mathrm{m}$. 
%This provides a natural cutoff for the model.

\begin{figure}[htbp]
    \centering  \includegraphics[width=0.8\linewidth]{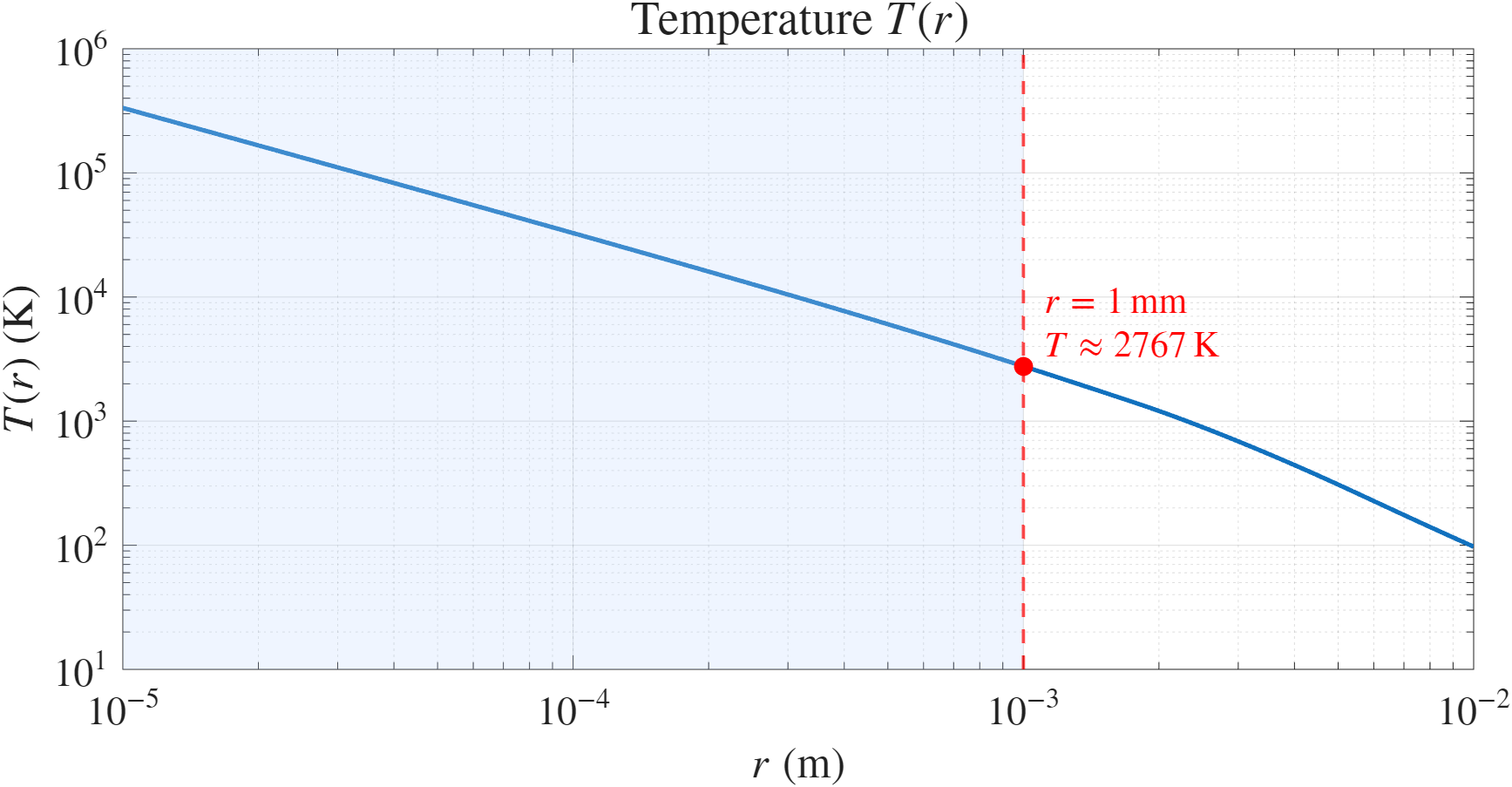}
    \caption{Calculated radial temperature profile in 316L stainless steel immediately after the passage of an AQN, obtained from the analytical energy-deposition model by solving Eq.~(\ref{Eq:epsilon(r)}) using temperature-dependent material properties. The temperature rises sharply toward the AQN trajectory because of the highly localized energy deposition. The red marker indicates the temperature at $r=1$ mm, where the material is heated to approximately $2.8\times10^3$ K. The blue region indicates radius scale smaller than $r=1$ mm where expected temperature is higher than that temperature.}
    \label{Fig:Radial_T_profile} 
\end{figure}

Equation \eqref{Eq:epsilon(r)}  was solved numerically using a C program that incorporates temperature-dependent 316L stainless steel properties. The resulting radial temperature increases towards the cut-off radius of $0.1\,\mu$m, see Fig.\,\ref{Fig:Radial_T_profile}. 
At $r \approx 0.1\,\mu\mathrm{m}$, the temperature exceeds $10^7$\,K. In this regime, the material is expected to enter a plasma state, and the model used for the calculation is no longer a good approximation. Even at macroscopic distances, the heating remains significant: at $r \sim 1\,\mathrm{mm}$, the temperature still reaches values on the order of $3000$\,K. This temperature, however, is above the melting temperature. 

\begin{figure}[htbp]
    \centering
    \includegraphics[width=0.8\linewidth]{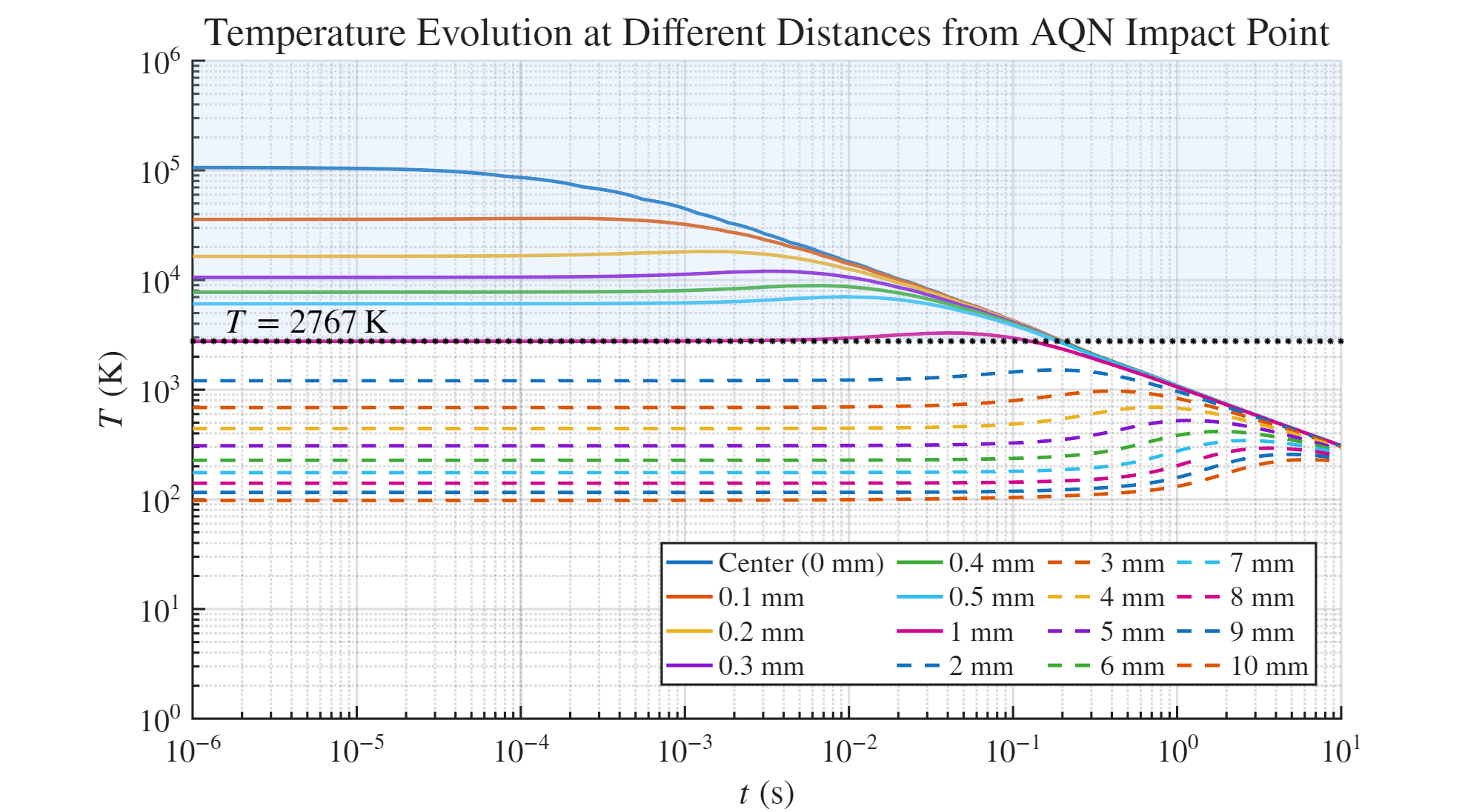}
    \caption{Time evolution of the temperature at different radial distances from the AQN trajectory obtained from the COMSOL finite-element simulation. Solid curves correspond to locations within 1 mm of the AQN path, whereas dashed curves represent larger distances. The simulation starts from the analytical radial temperature distribution shown in Fig.~\ref{Fig:Radial_T_profile}. The horizontal dotted line marks $T=2767$ K, corresponding to the starting temperature in $r=1$ mm. The blue region indicates temperature higher than that starting temperature.}
    \label{fig:Tt_profiles}
\end{figure}

To evaluate the dynamical consequences, not only the peak temperature but also its temporal evolution must be considered. For this purpose, a two-dimensional finite-element simulation was performed with COMSOL. The resulting time 
evolution of the temperature at different radial positions is shown in 
Fig.~\ref{fig:Tt_profiles}.

% \textcolor{red}{Zizhou, please make some of the lines dashed as several colors appear twice.}

A square domain of side length $0.05\,\mathrm{m}$ was used, discretized with a minimum mesh size of $0.05\,\mathrm{mm}$. The initial radial temperature distribution obtained from the analytical model was imposed as the initial condition.
The results show that the central temperature decreases rapidly due to heat conduction. Surrounding points always remain cooler than the center, though they may experience a transient temperature increase as heat flows outward.

Quantitatively, for the central region with a scale of 0.05\,mm:
\begin{itemize}
    \item $T > 10^4$\,K persists for approximately $10^{-2}$\,s,
    \item $T > 10^3$\,K persists for approximately $1$\,s.
\end{itemize}

Within $r \gtrsim 0.5\,\mathrm{mm}$ from the central AQN path, the temperature remains below $10^4$\,K at all times, and within $r \gtrsim 3\,\mathrm{mm}$ it remains below $10^3$\,K after the initial redistribution phase.

The results indicate that, within a time scale of order one second, regions extending to the millimeter scale may still experience extreme temperature. 
Such heating will have consequences for the material, such as evaporation, melting, structural damage, or may even leave an empty cylinder  along the AQN path.

\section{Discussions/Conclusions}

% \textcolor{red}{Let us discuss that, although we focused on the specific example of the direct detection of AQNs, in principle, the SF He-4 detector can also be used for detection of other rare events and DM candidates.}
In this work, we started by exploring the idea of using thermometry of superfluid helium in the LHC ring as a possible detection method for the rare AQN-passage events. However, in the course of the analysis, it became clear that the interaction of the AQN with the metallic enclosure will have significantly stronger, potentially spectacular signatures. The fact that no such events have apparently occurred in the nearly 20 years of the existence of the LHC sets, given the geometry of the machine, an upper limit of $\sim$3\,events/km$^2$/yr. This can be compared with the estimated AQN flux of $\sim0.04$($10^{25}/B)$ events/km$^2$/yr \cite{zioutas2025aqn,Budker_2020}.

% \DB{We find ourselves in a strange place where we might criticize the earlier critique of the papers that talk about limits from people as walking AQN detectors, the analysis of holes in bridges and building, etc. Another one: airplanes are super good detectors as well!}

% \medskip
% \DB{The rest of the text is from the older version before we realized the catastrophic effects of propagation in the metal.}
Although we focus on the specific example of the direct detection of DM AQNs, in principle, the SF He-4 detector can also be used for the direct detection of other rare events and DM candidates. To detect rare events from the dark sector within reasonable time interval, one is forced to increase the DM detector mass. Here we propose to use parasitically the LHC 27\,km cryo-ring as SF He-4 detector. This can be achieved without interfering with the LHC performance or the tight physics schedule of this worldwide unique collider and retroactively using the several years of the operation records.

The use of SF helium as DM detector is not new. However, the use of $\sim 85$\,t of superfluid helium-4 of LHC, along with the sensors distributed across the 27\,km LHC circumference, makes it a genuinely novel, large-scale segmented detector for rare events such as those expected from the dark sector. 

The primary motivation for this proposal is the detection of axion antiquark nuggets (AQNs), as proposed in \cite{Zhitnitsky:2002qa}. AQNs are expected to be macroscopic objects with $\sim 0.1\,\upmu$m size and nuclear density. The number of antiquarks inside the AQN created during early cosmological times is in the range of $10^{26\pm2}$. The estimated flux is $\sim0.04$($10^{25}/B)$ events/km$^2$/yr \cite{zioutas2025aqn,Budker_2020}. The exposure surface area of the SF He-4 is about 25 km $\times$ 60 cm $\approx$ 0.015\,km$^2$ which means that one can expect an AQN event during several years. In addition, the continuously stored LHC machine data cover the entire LHC operation since its start in 2011, with an integrated exposure time of $\sim 7$ yr.

Of note, with these two values for the sensitive surface area and the already existing data-taken period, the direct detection of AQNs using LHC machine data becomes possible, considering the naturally expected uncertainties in the underlying cosmological considerations that lead to the aforementioned flux estimation.
A significant additional boost in the event rate may occur due to the gravitational focusing of cosmic DM streams or clusters. Such objects were proposed in the past \cite{10.1111/j.1365-2966.2011.18224.x,TKACHEV1991289,PhysRevLett.71.3051}. For the widely assumed velocities of DM constituents, solar system planets act as gravitational lenses for DM streams. Gravitational focusing by solar system bodies (including the Moon) occurring occasionally can result in strong spatiotemporal flux enhancements \cite{PhysRevD.108.123043,Hoffmann:2003zz,Patla_2013,galaxies8020042,Prézeau_2015,PhysRevD.108.083001,bertolucci2017sunplanetsdetectorsinvisible,sym17010079}. Accounting for such enhancement of the event rate further boosts the discovery potential of our proposed method \cite{Rich}. 
If the event rate turns out to be sufficiently enhanced so that we observe multiple events, their temporal sequence can be analyzed to look for correlation with a  planetary periodicity. A temporal  alignment with a DM stream would stress the signature due to  dark matter streams being focused by the solar system bodies, while being aligned occasionally  with an incident stream \cite{PhysRevD.108.123043,Hoffmann:2003zz,Patla_2013,galaxies8020042,Prézeau_2015,PhysRevD.108.083001,bertolucci2017sunplanetsdetectorsinvisible,sym17010079}.

Another complimentary approach to search for the AQN signal using the LHC machine as the DM detector was recently proposed by \citet{Liang:2026tjs}. Then basic idea  is as follows. Since 2010, it has been recognized that the so-called Unidentified Falling Objects (UFOs) may pose a significant limitation to LHC performance as discussed in \cite{zioutas2025aqn}. UFO-LHC events  are commonly attributed to micrometer-sized dust particles released from the beam screen that become attracted to the proton beams and produce beam losses via inelastic proton-nucleus collisions. However, the mechanism that releases these dust particles remains an open question.   It has been suggested in \cite{Liang:2026tjs} that an  AQN crossing the atmosphere  100\,km away from the LHC area generates acoustic waves as described in \cite{Budker_2022} to trigger multiple UFO-LHC events.

% In fact, the perspective is to recover in the data analysis constant with time single planetary orbital periodicities, or synods of 2 or 3 planets. For example, the otherwise still mysterious 11 years solar cycle is identical to the synodic periodicity of Jupiter, Earth and Venus.

% In short, the already observed planetary periodicities with diverse solar or terrestrial observables mimic a not extant remote planetary force. A viable explanation, presented already, lies in the gravitational focusing of DM streams with focal regions within the solar system. Therefore, by observing a specific planetary periodicity one has at hand a new means for an unambiguous signal identification. We recall that expected typical solar periodicities cover about 4 orders of magnitude, between the ubiquitous sidereal daily and the 11 years solar cycle (including its 22 years rhythm).

% \textcolor{red}{KZ TBAdded on:}
% A FOURIER analysis of all data has the potential to derive characteristic periodicities resembling planetary dependencies. >> a more elaborated analysis projecting the time stamp of a solar observable allows to scrutiny possibly solar system hidden periodicities, and which do not appear some times in the FOURIER analysis.

%%%%%%%%%%%%%%%%%%%%%%%%%%%%%%%%%%%%%%%%%%
\authorcontributions{Conceptualization, K.Z. and all authors, simulations, M.P. and Z.H. All authors have written and edited the paper, read and agreed to the published version of the manuscript.}

% For research articles with several authors, a short paragraph specifying their individual contributions must be provided. The following statements should be used ``Conceptualization, X.X. and Y.Y.; methodology, X.X.; software, X.X.; validation, X.X., Y.Y. and Z.Z.; formal analysis, X.X.; investigation, X.X.; resources, X.X.; data curation, X.X.; writing---original draft preparation, X.X.; writing---review and editing, X.X.; visualization, X.X.; supervision, X.X.; project administration, X.X.; funding acquisition, Y.Y. All authors have read and agreed to the published version of the manuscript.'', please turn to the  \href{http://img.mdpi.org/data/contributor-role-instruction.pdf}{CRediT taxonomy} for the term explanation. Authorship must be limited to those who have contributed substantially to the work~reported.}

\funding{The work of M.P. is supported by the contract CERN-RO/CDI/$2024\_001$. The work of D.B. and Z.H. was supported by the DFG Project ID 390831469: EXC 2118 (PRISMA++ Cluster of Excellence), and by the COST Action within the project COSMIC WISPers (Grant No. CA21106). The stay of Z.H. in Mainz was supported by the USTC internship abroad program. A.Z. is supported in part by  the Natural Sciences and Engineering Research Council of Canada. }

\acknowledgments{The authors are indebted to G.\,Tsiledakis, and A.\,Wickenbrock for useful discussions and their help with this project.}

\conflictsofinterest{The authors declare no conflicts of interest. 
% \DB{Need to check citations, for example, \cite{Lebrun_SFHE-4} is incomplete and the title is ALL CAPS. Also, remember to get a sentence about airplanes and also to clean up LaTeX errors}
} 

%%%%%%%%%%%%%%%%%%%%%%%%%%%%%%%%%%%%%%%%%%

%\isPreprints{}{% This command is only used for ``preprints''.
%\begin{adjustwidth}{-\extralength}{0cm}

\reftitle{References}
\bibliography{References2_claude}

\PublishersNote{}
%\end{adjustwidth}
\end{document}